\documentclass[twocolumn,prl,aps,showpacs]{revtex4-1}
\usepackage{graphicx}
\usepackage{color}
\usepackage[T1]{fontenc}
\usepackage{bm}
\usepackage[utf8]{inputenc}
\usepackage{float}
\graphicspath{ {images/} }
\usepackage{amsmath}
\usepackage{amsfonts}

\usepackage[normalem]{ulem}

\begin{document}
	
	\title{Universality in ultradilute liquid Bose-Bose mixtures}
	\author{V. Cikojevi\'c, L. Vranje\v{s} Marki\'c}
	\affiliation{Faculty of Science, University of Split, Ru\dj era Bo\v{s}kovi\'ca 33, HR-21000 Split, Croatia}
	\author{G. E. Astrakharchik, J. Boronat}
	\affiliation{Departament de F\'{\i}sica, Universitat Polit\`ecnica de Catalunya,
		Campus Nord B4-B5, E-08034 Barcelona, Spain}
	
	\begin{abstract}
		We have studied dilute Bose-Bose mixtures of atoms with attractive
interspecies and repulsive intraspecies interactions using quantum Monte Carlo 
methods at $T=0$. Using a number of models for interactions, we determine the 
range of validity of the universal equation of state of the symmetric liquid 
mixture as a function of two parameters: the $s$-wave scattering length and the 
effective range of the interaction potential. It is shown that the 
Lee-Huang-Yang correction is sufficient only for extremely dilute liquids with 
the additional restriction that the range of the potential is small enough. 
Based on the quantum Monte Carlo equation of state we develop a new density 
functional which goes beyond the Lee-Huang-Yang term and use it together with 
local density approximation to determine density profiles of realistic 
self-bound drops.
	\end{abstract}
	
	\date{\today}
	
	\maketitle
	
	Dilute Bose and Fermi gases have proved to be a versatile tool for 
exploration of different phases of condensed-matter systems. For more than two 
decades, most of the experiments were done in the low-density gas phase, in the 
universal regime fixed solely by the  gas parameter $\rho a^3$, with $a$ the 
$s$-wave scattering length and $\rho$ the density. The range of universality of 
the homogeneous Bose gas was established using different model potentials and 
solving the $N$-body problem in an exact way with quantum Monte Carlo (QMC) 
methods~\cite{Giorgini99}. One of the most important advances in the field of 
ultracold atoms in the last years is the recent creation of ultradilute quantum 
droplets. Such self-bound quantum systems were first experimentally observed for 
dipolar atoms~\cite{kadau,ferrier,schmitt,schauss} being caused by a close 
cancellation of the dipolar and short-range energies. 
Petrov~\cite{petrov} pointed out that liquid drops can be created in an even 
simpler setup composed by a two-component mixture of bosons with short-ranged 
attractive interspecies and repulsive intraspecies interactions. However, the 
perturbative technique employed by Petrov is valid only very close to the 
mean-field (MF) instability limit, that is for extremely dilute liquids. The 
collapse predicted on the MF level is avoided by stabilization due 
to the quantum fluctuations described by the Lee-Huang-Yang (LHY) correction to 
the energy. It was shown that a similar stabilization mechanism can be used in 
two- and one-dimensional geometries where the resulting liquid phase has 
enhanced stability~\cite{petrov2}. Very recently, two experimental groups 
managed to obtain self-bound liquid drops~\cite{tarruell,semeghini} which, upon 
releasing the trap, did not expand. The drops required a certain critical number 
of atoms to be bound. Importantly, measurements of the critical number and size 
of the smallest droplets could not be fully accounted for by the MF+LHY 
term~\cite{tarruell}.
	
	Recently, some of us have studied liquid Bose-Bose droplets by using 
	the diffusion Monte Carlo (DMC) method, thus solving \textit{exactly} the 
full many-body problem for a given Hamiltonian at zero 
temperature~\cite{cikojevic}. Our results have confirmed the transition from a 
gas, with positive energy, to a self-bound droplet with negative energy. 
Furthermore, we have determined the critical number of atoms needed to form a 
liquid droplet as a function of the intraspecies scattering length. Using two 
different models for the attractive interaction, we did not get quantitatively 
the same results for the range of scattering lengths studied, which points to 
the lack of universality in terms of $\rho a^3$. It is thus of a fundamental 
interest to find whether there is a range of densities and scattering lengths 
where such universality exists. This is in fact expected when the system is very 
close to the MF collapse. In the case of homogeneous Bose gases, departures 
from universality start to appear around $\rho a^3 \gtrsim 
10^{-3}$~\cite{Giorgini99}. In that case, adding the LHY correction allowed for 
a good approximation of the equation of state up to higher densities. Recently, 
a variational hypernetted-chain Euler-Lagrange calculation~\cite{staudinger} of 
unbalanced mixtures showed that the drops can only be stable in a very narrow 
range,	near an optimal ratio of partial densities and near the energy minimum. 
Moreover, Ref.~\cite{staudinger} found dependence on the effective range even at 
low densities.
	
	In this Letter, we use the DMC method to address the question of the 
universality in the equation of state of dilute Bose-Bose mixtures. The second 
question 
we pose here is whether  there exists a regime where instead of using only one 
parameter ($s$-wave scattering length) inclusion of an additional parameter 
(effective range) extends the validity of the universal description. To answer 
these questions directly for finite-size droplets would require enormous 
computational resources, as at least thousands of atoms are needed to achieve a 
self-bound state close to the mean-field limit~\cite{tarruell}. In order to 
eliminate the finite-size effects caused by the surface tension and simplify the 
analysis, we study here bulk properties corresponding to the interior of  
large saturated droplets. From the obtained equation of state we construct a new 
density functional and use it to predict the profiles of the drops, discussing 
the effects of the potential range. 

We rely on the DMC method, which was  
successfully used in the past for determining the ground-state properties of 
interacting many-body systems.
	The DMC method stochastically solves the imaginary-time Schr\"odinger 
equation, giving for bosonic systems exact results within the statistical 
noise~\cite{casu1}. The Hamiltonian of our system is given by
	\begin{equation}
	H = -\sum_{\alpha=1}^{2}\frac{\hbar^2}{2 m_{\alpha}} \sum_{i=1}^{N_{\alpha}} \nabla_{i\alpha}^2
	+\frac{1}{2}\sum_{\alpha, \beta=1}^{2} \sum_{i_{\alpha},
		j_{\beta=1}}^{N_{\alpha}, N_{\beta}} V^{(\alpha, \beta)}
	(r_{i_{\alpha}j_{\beta}}) \ ,
	\label{hamiltonian}
	\end{equation}
	where $V^{(\alpha,\beta)}(r)$ is the interatomic interaction between species $\alpha$ and $\beta$. The intraspecies interactions with positive $s$-wave scattering length are modeled either by a hard-core potential of diameter $a_{ii}$ or by a 10-6 potential~\cite{10-6_exact} that does not support a two-body bound state,
	$V(r) =  V_0 \left[ \left(\frac{r_0}{r}\right)^{10} -
	\left(\frac{r_0}{r}\right)^{6} \right] $.
	The latter model has an analytic scattering length given in Ref.~\cite{10-6_exact}. The interspecies interactions with negative scattering length, $a_{12} < 0$, are modeled by a square-well potential of range $R$ and depth $-V_0$ or by a 10-6 potential with no bound states.
	We resort to a second-order DMC method and use a guiding wave function to reduce the variance, as described in Ref.~\cite{casu1}. The trial wave function is constructed as a product of Jastrow factors~\cite{suplem,reatto}.
	
	We consider a mixture with equal masses of particles $m_1=m_2=m$. Such situation is typical in experiments where different hyperfine states of the same atomic species are used to create two components~\cite{tarruell}. Furthermore, in order to reduce the number of degrees of freedom we choose to study the symmetric mixture with $a_{11}=a_{22}$ resulting in $N_1=N_2$. 
	The calculations are performed in a box with periodic boundary conditions. 
	We carefully analyze the finite-size effects, as discussed in 
Ref.~\cite{suplem}. We have also optimized the time-step and 
population bias to reduce their influence below the statistical noise.
	
	
	
	\begin{figure}[!htb]
		\centering		
		\includegraphics[width=7cm]{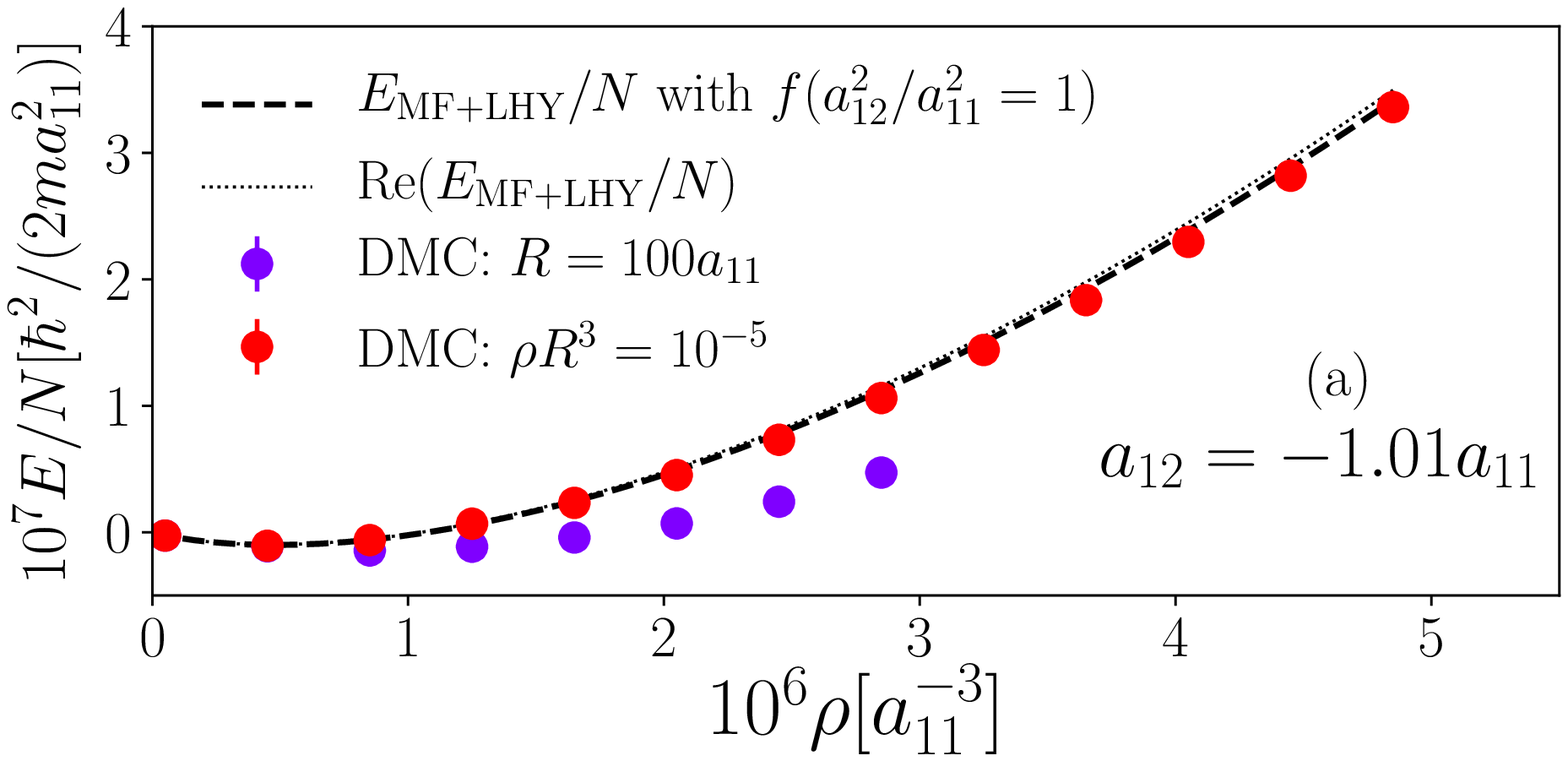}
		\includegraphics[width=7cm]{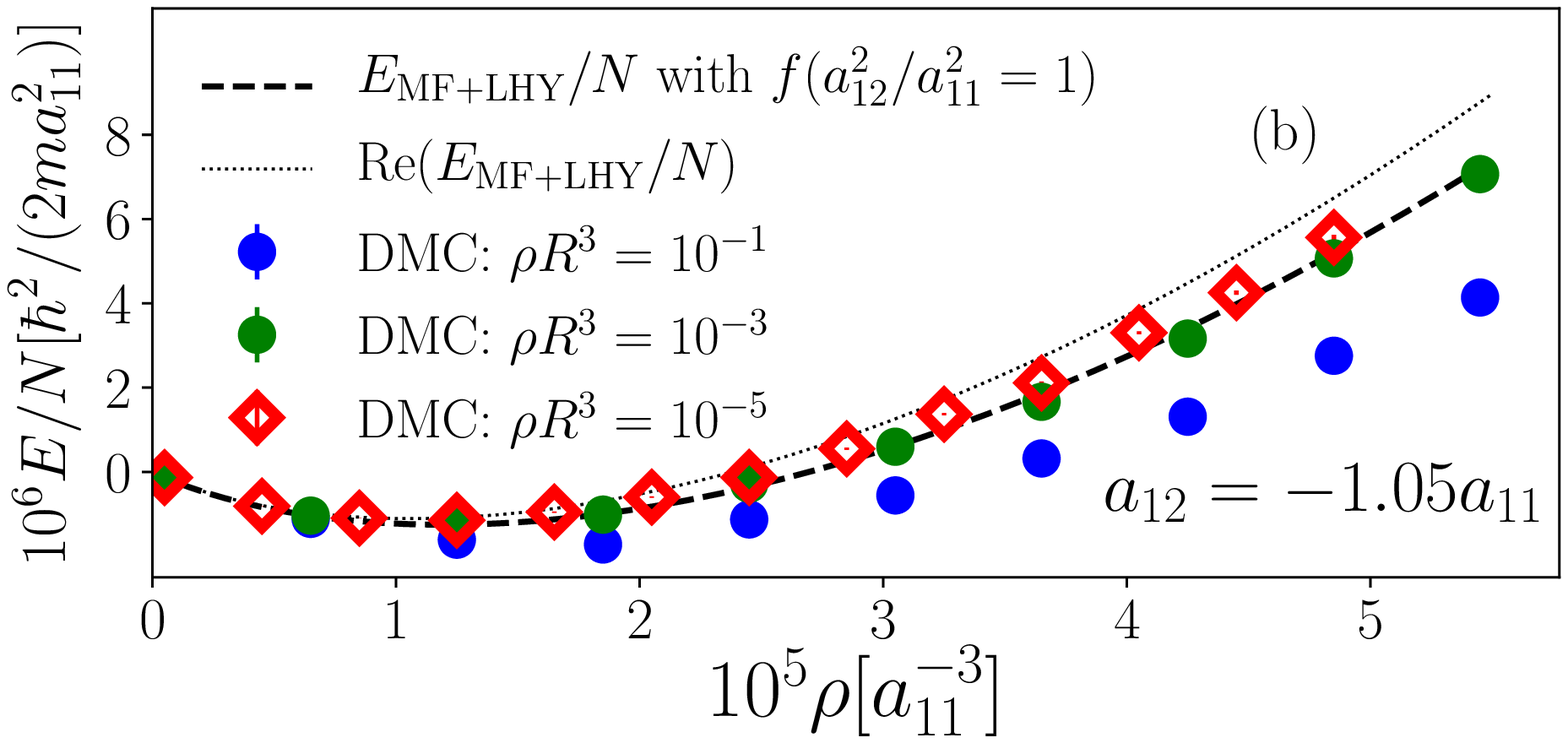}
		\includegraphics[width=7cm]{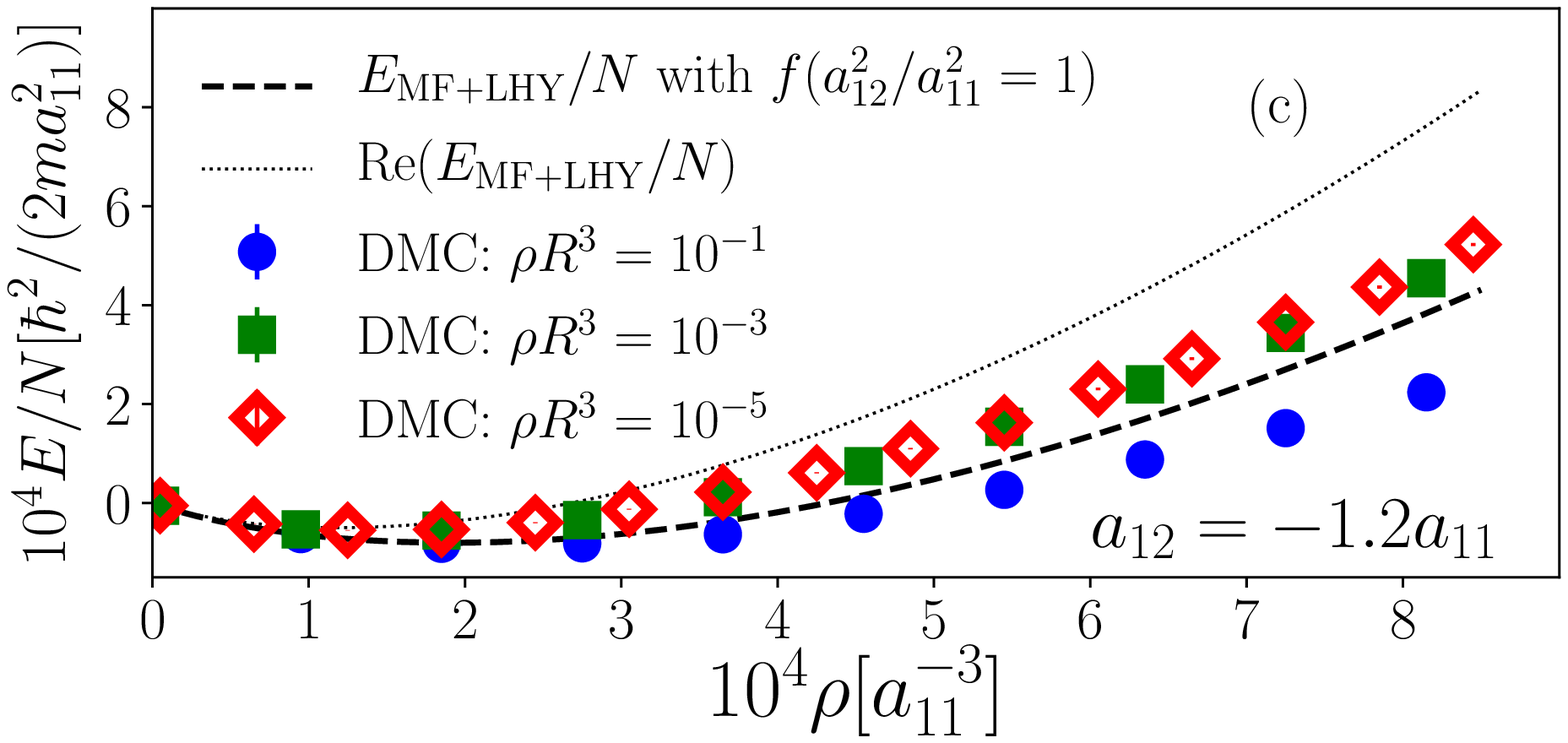}
		\caption{DMC equation of state of the liquid mixture for different
			values of $a_{12}$ and different ranges $R$, in comparison with MF+LHY
			theory.}
		\label{fig:fig1}
	\end{figure}
	First, we report results obtained using the hard-core model for the 
repulsive interactions and a square-well (SW) potential for the attractive 
ones. 
In Fig.~\ref{fig:fig1}, we show our results for different values of the 
interspecies scattering length $a_{12}$ and different ranges of the attractive 
well $R$, and compare them to the MF+LHY prediction~\cite{petrov}. The equation 
of state in Ref.~\cite{petrov} for $m_2 = m_1=m$, $a_{22} = a_{11}$, and $N_2 = 
N_1$ is given by		
	\begin{equation}
	\frac{E}{N}
	= \frac{\hbar^2 \pi (a_{11} + a_{12})}{m} \rho
	+
	\frac{32\sqrt{2 \pi}}{15}\frac{\hbar^2 a_{11}^{5/2} }{m} f\left(\frac{a_{12}^2}{a_{11}^2}\right) \rho^{3/2}\ ,
	\label{eq:petrov}
	\end{equation}
	with $f(x) = (1 + \sqrt{x})^{5/2} + (1- \sqrt{x})^{5/2}$.
	Notice that the function $f(x)$ becomes complex for $a_{12} < -a_{11}$
	and the presence of the imaginary component reduces the applicability of 
the perturbative theory. If instead the argument is approximated by $x = 
a_{12}^2 / a_{11}^2 = 1$ so that the function $f(x)$ remains real, as it was 
done in Ref.~\cite{petrov}, Eq.~(\ref{eq:petrov}) reduces to the following 
form,	
	\begin{equation}
	\frac{E}{N}
	= \frac{\hbar^2 \pi (a_{11} + a_{12})}{m} \rho
	+
	\frac{256\sqrt{\pi}}{15}\frac{\hbar^2 a_{11}^{5/2} }{m}  \rho^{3/2}	\ .
	\label{eq:petrovs}
	\end{equation}
	shown with a dashed line in Fig.~\ref{fig:fig1}. We plot as well  the 
energy resulting from taking the real part of $f(x)$~(\ref{eq:petrov}), without 
invoking the approximation $x = 1$. Only very close to the $a_{12}=-a_{11}$ 
limit corresponding to zero equilibrium density, both predictions are nearly the same while 
they clearly differ for finite densities. We report the exact DMC energies in 
Fig.~\ref{fig:fig1}. The perturbative MF+LHY results are recovered for small 
range $R$ of the square well and $\rho a_{11}^3 \approx 10^{-6}$, see 
Fig.~\ref{fig:fig1}a. However, when $R$ is increased by a large amount (to 
$R=100 a_{11}$) the universality breaks at $\rho R^3 \simeq 10^{-1}$. The 
energies for experimentally relevant densities, $\rho a_{11}^3 \approx 10^{-5}$ 
~\cite{tarruell,semeghini}, are reported in Fig.~\ref{fig:fig1}b.
	In this case and for larger densities (Fig.~\ref{fig:fig1}c), we observe 
that the energy depends on the potential range. Furthermore, the two ways of 
writing the perturbative equation of state, given by 
Eqs.~(\ref{eq:petrov},~\ref{eq:petrovs}), differ among themselves but are not 
equal to the obtained DMC equation of state. The latter appears to be 
independent of $R$ up to approximately $\rho R^{3}=10^{-3}$. Indeed, the 
difference between the energy per particle $E/N$ calculated at $\rho 
R^{3}=10^{-3}$ and $\rho R^{3}=10^{-5}$ is at most 3 errorbars, or 6\% at the 
highest density and at most 4\% in the minimum. 	
	
	It can be noted that within perturbative theory the energy is a single 
curve written in units of the equilibrium energy $E_0$ and density $\rho_0$. 
That is, the equation of state~(\ref{eq:petrovs}) can be conveniently 
represented as a $(E / E_0, \rho / \rho_0)$ curve,
	\begin{equation}
	\dfrac{E}{|E_0|} =
	-3 \left(\frac{\rho}{\rho_0}\right)
	+2 \left(\frac{\rho}{\rho_0}\right)^{3/2}\ ,
	\label{eq:petrovs2}
	\end{equation}	
		with, for the symmetric mixture, $\rho_0 = 25 \pi  \left(a_{11} + 
a_{12}\right)^2/(16384 a_{11}^5)$ and $E_0/N = -25 \pi^2 \hbar^2 |a_{11} + 
a_{12}|^3 / (49152 m a_{11}^5)$.	%
	\begin{figure}[tb]
		\begin{center}
			\includegraphics[width=0.9\linewidth]{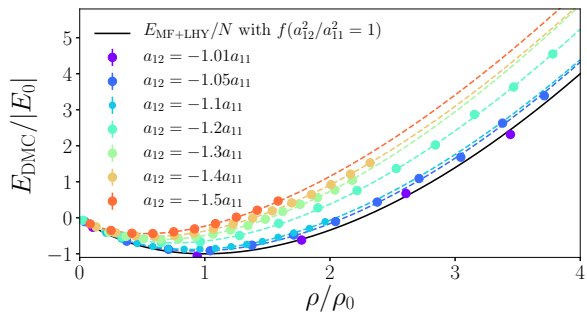}
			\caption{Equations of state for different $a_{12}/a_{11}$ normalized to the density and the energy at MF+LHY equilibrium point $(\rho_0,E_0)$. Dashed lines show fits to the data in the form of $E/N = \alpha x + \beta x^{\gamma}$ with $x=\rho a_{11}^3$. The range of the square well is $\rho R^3 = 10^{-5}$.}
			\label{fig:fig2}
		\end{center}
	\end{figure}	
	
	The DMC equations of state for different scattering lengths are shown in 
Fig.~\ref{fig:fig2}. The results are obtained for sufficiently small potential 
range, $\rho R^{3}=10^{-5}$, ensuring the universality in terms of the $s$-wave 
scattering length. As already observed in Fig.~\ref{fig:fig1}, when $|a_{12}| 
\approx a_{11}$, the MF+LHY prediction is recovered. Increasing 
$|a_{12}|/a_{11}$ repulsive contributions to the energy beyond the LHY terms are 
found. 
At the same time, the equilibrium densities become lower compared to the ones 
predicted by Eq.~(\ref{eq:petrovs2}), which was obtained by calculating $f(x)$ 
function at $x=1$. If instead one uses Eq.~(\ref{eq:petrov}) derived by taking 
the real part of $f(x)$, weaker binding is predicted as compared to DMC results. 
Thus, as we can see in Fig.~\ref{fig:fig1}, for small ranges $\rho 
R^{3}=10^{-5}$, the DMC many-body prediction is between Eq.~(\ref{eq:petrov}) 
and (\ref{eq:petrovs2}), but closer to Eq.~(\ref{eq:petrovs2}).
	\begin{table}[tb]	
		\caption{Energies (equilibrium and spinodal) and densities for different scattering lengths $a_{12}/a_{11}$ for small ranges $\rho R^3 = 10^{-5}$. Here ``eq'' stands for the minimum from the fit to DMC energy shown in Fig.~\ref{fig:fig2}, ``0'' stands for minimum of perturbative equation of state given by Eq.~(\ref{eq:petrovs2}), spinodal point is denoted by ``sp'' from the fit on DMC data  and ``sp,0'' in case of Eq.~(\ref{eq:petrovs2}). 
		}
		\centering
		\resizebox{0.499\textwidth}{!}{	
			\begin{tabular}{c c  c c c c  c}	
				\hline \hline	\\					
				$~~\dfrac{a_{12}}{a_{11}}~~$	& $~\dfrac{10^5\rho_{\rm eq}}{a_{11}^{-3}}~$	& $~\dfrac{\rho_{\rm eq}}{\rho_{\rm 0}}~$	& $~\dfrac{10^5\rho_{\rm sp}}{a_{11}^{-3}}~$		& $~\dfrac{\rho_{\rm sp}}{\rho_{\rm sp,0}}~$	& $~\dfrac{10^6\hbar^2E_{\rm eq}}{2 m a_{11}^2N}   ~~$	&  $~~\dfrac{E_{\rm eq}}{E_{\rm 0}}~ $	\\
				\hline \\		
				$-1.05	$&$1.12	$&$0.934	$&$0.715	$&$0.932	$&$-1.15	$&$0.919$\\
				$-1.10	$&$4.28	$&$0.894	$&$2.73 	$&$0.888	$&$-8.82	$&$0.879$\\
				$-1.20	$&$14.5	$&$0.754	$&$9.19 	$&$0.749	$&$-56.0	$&$0.697$\\
				$-1.30	$&$28.0	$&$0.649	$&$17.7 	$&$0.641	$&$-163	$&$0.601$\\
				$-1.40	$&$44.9	$&$0.585	$&$28.3     $&$0.576	$&$-334		$&$0.520$\\
				$-1.50	$&$62.4	$&$0.521	$&$39.3     $&$0.512	$&$-554	$&$0.441$\\
				\hline	
				\label{tab:en}						
		\end{tabular}	}										
	\end{table}
	The DMC values of the equilibrium energies and densities are reported in 
Table~\ref{tab:en}. They are also compared to predictions from perturbative 
theory given by Eq.~(\ref{eq:petrovs2}). With the increase of $|a_{12}|/a_{11}$ 
the equilibrium and spinodal densities start to depart significantly from the 
MF+LHY values. It is worth noticing again that the MF+LHY equation of state 
becomes complex, and thus unphysical, unless the approximation $f(a_{12}^2 / 
a_{11}^2 = 1)$ is used. Our results show that, even very small (in absolute 
value) negative pressures, can cause spinodal instability. For typical 
experimental parameters $a_{11}=50a_0$~\cite{tarruell,semeghini} the uniform 
liquid breaks into droplets when the applied negative pressure is very small, 
from $1.81$pPa for $a_{12}=-1.05 a_{11}$ to $31.3$nPa for $a_{12}=-1.5 a_{11}$.
	
\begin{figure}[tb]
		\centering
		\includegraphics[width=0.9\linewidth]{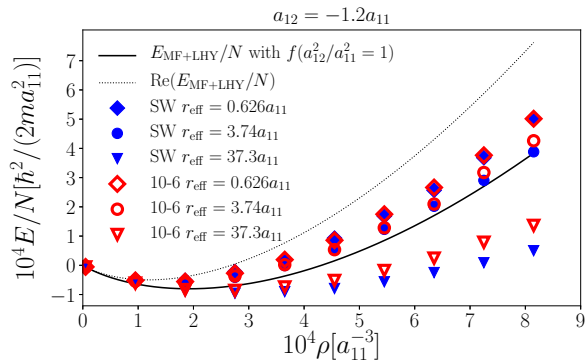}
		\caption{Dependence of the equation of state on the effective range.}
		\label{fig:fig3}
	\end{figure}

	As can be seen from Fig.~\ref{fig:fig1}, the equation of state loses 
universality in terms of the scattering length when $\rho R^3 \gtrsim 10^{-3}$. 
This poses  the relevant question if whether by fixing one more parameter, 
besides the $s$-wave scattering length, it is possible to obtain a universal 
description. 
	To address this question, we performed DMC calculations using the 10-6 model
with equivalent values of the $s$-wave scattering lengths and effective range 
$r_{\rm eff}$ of the attractive interaction. For the repulsive interactions, we 
fix the range of the 10-6 model potential to $r_0 = 2a_{11}$. In 
Fig.~\ref{fig:fig3}, we show results for scattering length $a_{12}=-1.2a_{11}$ 
and three values of the effective range $r_{\rm eff}$. The solid line is for 
Eq.~(\ref{eq:petrovs}) and the dashed one for the real part of 
Eq.~(\ref{eq:petrov}). The range of the SW potential is $R/a_{11}$ = 0.531, 
2.17, and 9.18 when $r_{\rm eff} / a_{11} = 0.626$, 3.74 and 37.3, respectively. 
We find that, specifying only the scattering length, one cannot generally 
obtain universal results unless the range is sufficiently small, $\rho 
R^3 \lesssim 10^{-1}$. The interaction potential for a given  scattering length 
predicts different energies and equilibrium densities when different effective 
ranges are used. Generally, increasing the range lowers the energy and shifts 
the equilibrium density to larger values. However, if we specify both the 
scattering length and the effective range, then we observe that the difference 
between results of two models is always smaller than the difference between 
results for the same type of model but with different range. In 
Fig.~\ref{fig:fig3}, the two models with $r_{\rm eff}/a_{11}=0.626$ give, within 
errorbars, the same energies in the whole density range. Increasing the range, 
at higher densities we observe that the two potentials start to give different 
predictions and that the difference between them grows with the increase in 
density. Interestingly, even when the effective range is quite large, $r_{\rm 
eff}/a_{11} = 33.6$, the relative difference between the models remains lower 
than 10\%, as long as $\rho R^3 < 0.2$. Increasing the density even further,   
we would need more parameters beyond $a_{12}$ and $r_{\rm eff}$ to describe the 
interaction.
	The observed dependence on the effective range for $\rho_{eq} R^3>10^{-3}$ 
is in overall agreement with recent calculation of unbalanced 
mixtures~\cite{staudinger} based 
on the variational HNC method. It is interesting to notice that the MF+LHY 
equations of state, following Eq.~(\ref{eq:petrovs2}), are actually closer to 
our full many-body calculations using rather large values of the effective 
range. On the other hand, the results using only the real part of 
Eq.~(\ref{eq:petrov}) are above the DMC energies for even the smallest range.
	
	Presuming that the equation of state of the liquid mixture is universal in terms of the scattering length and the effective range for $\rho R^3 \lesssim 10^{-1}$, we use the SW results to deduce the following form for the equation of state	
	\begin{equation}
	\dfrac{E}{N}=\dfrac{|E_0|}{N} 		
	\left[
	-3\left(\dfrac{\rho}{\rho_0}\right)
	+ \beta \left(\dfrac{\rho}{\rho_0}\right)^{\gamma}		
	\right] \ ,
	\label{eos}
	\end{equation}
	where $\beta$ and $\gamma$ are functions of $a_{12}/a_{11}$ and $r_{\rm eff}/a_{11}$(see Supplementary Materials~\cite{suplem} for the specific values).	
	The equation of state~(\ref{eos}) can be used as an energy 
functional~\cite{barranco} to calculate density profiles of liquid mixture drops 
within the local density approximation (LDA). 	The results for the equilibrium 
density as a function of the interspecies scattering length and the square-well 
range are presented in Fig.~\ref{fig:fig4} and compared to the MF+LHY 
predictions, while the full density profiles are given in the Supplementary 
information~\cite{suplem}. For a negligible range $R$, the equilibrium density 
drops below 
the MF+LHY prediction as $|a_{12}|/a_{11}$ is increased. The effect of the 
finite range is to increase the equilibrium density. That is by increasing $R$, 
the LDA prediction crosses the perturbative result of MF+LHY and goes above. 
Overall, by increasing the range and decreasing $|a_{12}|/a_{11}$ (i.e. going in 
the up-right direction in Fig.~\ref{fig:fig4}) we observe an increase of 
$\rho_{\rm LDA}^{\rm eq} / \rho_{\rm MF+LHY}^{\rm eq}$.
	\begin{figure}[tb]
		\centering
		\includegraphics[width=0.9\linewidth]{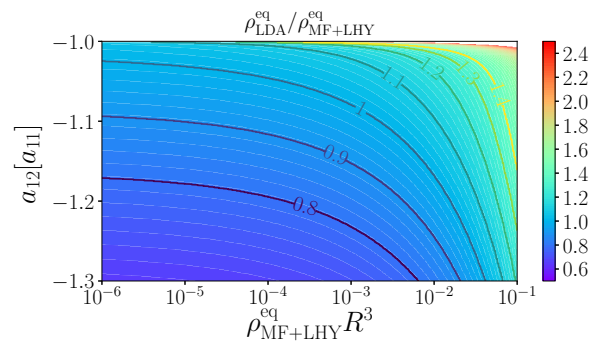}
		\caption{Ratio of equilibrium densities $\rho_{\rm LDA}^{\rm eq} / \rho_{\rm MF+LHY}^{\rm eq}$ vs. scattering length $a_{12}$ and the range $R$. Isolines follow values with the constant $\rho_{\rm LDA}^{\rm eq} / \rho_{\rm MF+LHY}^{\rm eq}$. LDA results are obtained starting from the DMC equation of state. }
		\label{fig:fig4}
	\end{figure}
	
	In conclusion, we have carried out high-precision DMC calculations of the 
ground-state equation of state of ultradilute two-component Bose liquids. 
We find out that the use of only the first beyond-MF correction, the LHY term, is 
accurate only for extremely small densities and only when the range of the 
interaction is not very large. In our study, we have used for the range $R$ the 
diameter of the square well potential, which has the same scattering length and 
effective range as the chosen model. If $|a_{12}/a_{11}+1| \le 0.05$ and $\rho R^3 < 
10^{-3}$ one parameter, the s-wave scattering length is enough to describe the 
system, but there is an appreciable difference with respect to MF+LHY. 
Increasing the range, one enters in a regime where interaction potentials with 
the same scattering length and effective range give equivalent results within 
10\%, which means that up to $\rho R^3 = 0.1$ we have at hand a  universal 
equation of state which is function of two parameters. For even larger values of 
$\rho R^3$  additional parameters would need to be specified. The results of 
scattering calculations of alkali atoms, such as given
	in Ref.~\cite{flambaum,tanzi}, indicate that most likely the effective 
	ranges are quite far from the zero-range limit. In that case, for obtaining 
the correct results one needs a full many-body approach like DMC. 
	Here, we provide a new energy functional based on the best fit to DMC data 
and use it to calculate the density profiles of realistic drops with LDA. 
	\acknowledgments
	We acknowledge fruitful discussions with Leticia Tarruell. This work has been supported in part by the Croatian Science Foundation under the project number IP-2014-09-2452. Partial financial support from the MINECO (Spain) grant No. FIS 2014-56257-C2-1-P and No. FIS2017-84114-C2-1-P are also acknowledged. The computational resources of the Isabella cluster at Zagreb University Computing Center (Srce) and Croatian National Grid Infrastructure (CRO NGI) were used. The Barcelona Supercomputing Center (The Spanish National Supercomputing Center - Centro Nacional de Supercomputaci\'on) is acknowledged for the provided computational facilities (RES-FI-2018-3-0027).

\end{document}